\documentclass{gistt}

\usepackage[capitalise]{cleveref}
\usepackage{graphicx}
\usepackage{subcaption}

\title{Should I Run My Cloud Benchmark on Black Friday?}

\author{%
    Sören Henning,\textsuperscript{1}
    Adriano Vogel,\textsuperscript{1}
    Esteban Perez-Wohlfeil,\textsuperscript{1}
    Otmar Ertl,\textsuperscript{1}
    Rick Rabiser\textsuperscript{2}\\%
    \textsuperscript{1}Dynatrace Research, Linz, Austria\\%
    \textsuperscript{2}LIT CPS Lab, Johannes Kepler University Linz, Austria\\%
    \{firstname.lastname\}@\{dynatrace.com, jku.at\}%
}

\begin{document}

\maketitle

\begin{abstract}
Benchmarks and performance experiments are frequently conducted in cloud environments.
However, their results are often treated with caution, as the presumed high variability of performance in the cloud raises concerns about reproducibility and credibility.
In a recent study, we empirically quantified the impact of this variability on benchmarking results by repeatedly executing a stream processing application benchmark at different times of the day over several months. Our analysis confirms that performance variability is indeed observable at the application level, although it is less pronounced than often assumed. The larger scale of our study compared to related work allowed us to identify subtle daily and weekly performance patterns.
We now extend this investigation by examining whether a major global event, such as Black Friday, affects the outcomes of performance benchmarks.
\end{abstract}

\section{Introduction}

With the ongoing transition to cloud-based deployments in many organizations, conducting benchmarks and performance experiments in the cloud has also become common practice in both research and engineering. Cloud environments are widely available in many organizations and provide a realistic testbed that closely reflects production deployments.
However, caution is urged when interpreting performance measurements in such settings~\cite{Papadopoulos2021}. Cloud workloads share underlying infrastructure with other tenants and the high abstraction from hardware introduces variability~\cite{Leitner2016,Abedi2017}.
As a result, performance measurements may fluctuate, impairing reproducibility and hindering meaningful comparisons across studies.

In a recent study~\cite{FSE2025}, we empirically assessed the significance of performance variability in the cloud through a longitudinal investigation for the case of distributed stream processing applications.
By repeatedly executing the same benchmark over several months, we collected a large dataset that enables quantitative analysis of performance fluctuations and, beyond related work,\footnote{See our conference paper for a more detailed discussion of the literature~\cite{FSE2025}.} provides an updated, application-level characterization over longer time spans.
In this paper, we summarize our key findings and complement them by an investigation of cloud performance variability in the context of the Black Friday event. Black Friday is a global shopping event that is known to cause massive increases in web traffic and, hence, cloud resource demand~\cite{Horwitz2022}. We investigate whether such an event measurably affects the observed performance of benchmark executions.

\section{Experiment Design}

\paragraph{Test subject.}
Our study design focuses on stream processing applications as a representative type of data-intensive, performance-critical distributed systems~\cite{SEAA2023}. These applications process continuous streams of data with low (often sub-second) latency, involve heavy CPU and network usage, while also having to maintain properties such as fault-tolerance, scalability, resource efficiency. As test subject for our experiments we use the open-source, cloud-native stream processing benchmark ShuffleBench~\cite{ICPE2024} with its Kafka Streams implementation.

\paragraph{Performance measurements.}
As performance metric for which we quantify variability, we focus on throughput captured according to ShuffleBench's ad-hoc measurement method~\cite{ICPE2024}. The obtained throughput values provide a good estimate of the load a similar real-world application could sustain under typical operating conditions. Achieving high throughput while minimizing required computing resources is a core optimization target. 
Within one benchmark execution, we continuously measure the application's throughput, discard the first 3~minutes as warm-up, and take the average over the remaining duration as result of the benchmark. Each benchmark execution thus produces one average throughput value for the subsequent analysis in \cref{sec:patterns} and \cref{sec:black-friday}.

\paragraph{Execution environment.}
We run our benchmarks in managed Kubernetes environments, which is a common choice for operating large-scale, distributed, and data-intensive software systems in the cloud.
We use the largest cloud provider Amazon Web Services (AWS) with its Elastic Kubernetes Service (EKS) offering. The Kubernetes cluster consists of 10~nodes provisioned with \textit{m6i} instances of different size~\cite{FSE2025} in the \textit{us-east-1} region.

\paragraph{Automated benchmark execution.}
A scheduled task in AWS Elastic Container Service (ECS) automates periodic benchmark execution. It provisions a new EKS cluster and installs the benchmarking infrastructure, including Apache Kafka, monitoring tooling, and the Theodolite benchmarking framework~\cite{EMSE2022}. Once the setup is complete, the ECS task initiates the execution of the ShuffleBench benchmark through Theodolite.
Theodolite launches the stream processing application and ShuffleBench's load generator, keeps them running for 15~minutes, and collects the monitored throughput data.
The benchmark execution is repeated three times according to Theodolite's configuration.
Finally, benchmark results are stored for later analysis, the benchmarking infrastructure is uninstalled, and the cluster is decommissioned.

\paragraph{Time spans.}

Between May and July 2024 as well as for one week in September 2024, the periodic benchmarking task was configured to run every 6 hours to cover a full daily cycle. For a period of 3 weeks, we additionally reduced the time between experiments to 3 hours to capture a more fine-grained daily pattern.
Each task execution runs the benchmark three times to account for performance variability within the same infrastructure.
In \cref{sec:black-friday}, we report on an additional week of experiments around Black Friday 2024.

\section{Cloud Performance Patterns}\label{sec:patterns}

\begin{figure}%
	\centering%
	\includegraphics[width=\linewidth]{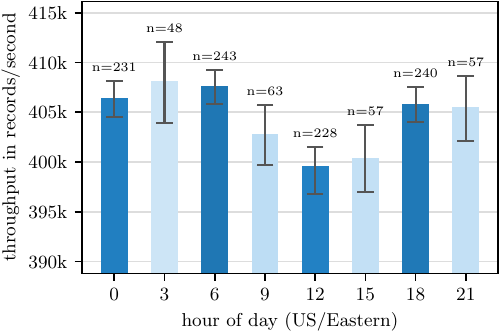}%
	\caption{Measured throughput summarized by the hour of the day; color intensity reflects sample size~\cite{FSE2025}.}%
	\label{fig:daily}%
\end{figure}

From our more than one thousand benchmark executions, we obtain a clear picture of performance variability in the cloud. The distribution of throughput measures shows a clear central tendency and almost symmetry in the interquartile range, but is not normal distributed due to slight skewness toward lower throughput results. 
We observe a coefficient of variation (CV), a common measure for quantifying performance variability, of 3.69\%. This is on the lower end of the wide range of
variability reported for micro and system-level benchmarks in the literature. Moreover, 50\% of all measurements are within $-2.4\%$ and $+2.3\%$ of the median (i.e., the interquartile range).
This leads us to conclude that cloud performance variability clearly exists, but contrary to what is sometimes assumed, it is not inherently detrimental
when benchmarking on the application level.

\paragraph{Daily pattern.}
To break down our results, we investigate whether the performance variability exposes a daily pattern. For this purpose, we summarize all results by the hour of the day when
the corresponding experiment was executed. \Cref{fig:daily} shows the mean observed
throughput per hour of day with its corresponding confidence intervals (obtained via bootstrapping).
We observe a subtle yet statistically significant daily pattern in performance. Benchmarks executed around noon tend to exhibit slightly lower performance, whereas those conducted during late-night and early-morning hours achieve the highest results with a difference of the mean of $2.15\%$.

\begin{figure}%
	\centering%
	\includegraphics[width=\linewidth]{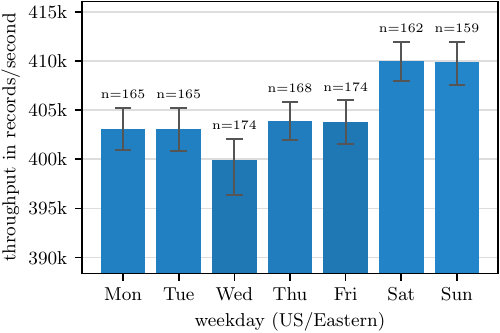}%
	\caption{Measured throughput summarized by the day of the week; color intensity reflects sample size~\cite{FSE2025}.}%
	\label{fig:weekly}%
\end{figure}

\paragraph{Weekly pattern.}
A similar analysis breaks down the observed throughput by the day of the week as shown in \cref{fig:weekly}. Again we observe a modest yet statistically significant pattern in performance. Benchmarks executed over the weekend show slightly higher performance compared to weekdays, with Wednesday standing out as the day with the lowest performance. The maximum variability is similar to the daily pattern with a difference of $2.52\%$ in mean throughput from Saturdays to Wednesday.

\paragraph{Long-term pattern.}
Breaking down the results by the week of execution reveals small performance fluctuations over time. However, from our result we cannot see a long-term pattern or trend. As our experiments span only part of the year, we cannot rule out the possibility of differences during other periods.

\paragraph{Further observations.}
Besides these findings for the \textit{us-east-1} cloud region with \textit{m6i} instances, we made similar observations in experiments conducted in the \textit{eu-central-1} region or with \textit{m6g} instances.

\section{Cloud Performance on Black Friday}\label{sec:black-friday}

\begin{figure}%
	\centering%
	\includegraphics[width=\linewidth]{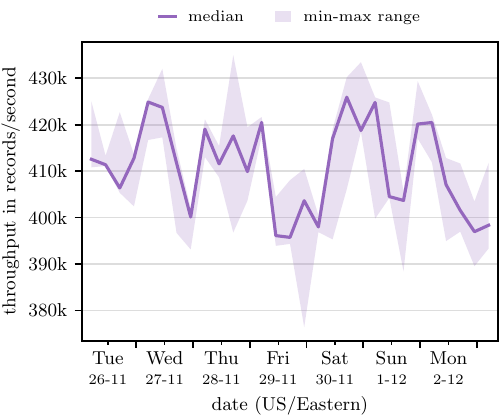}%
	\caption{Measured throughput in benchmark executions around Black Friday 2024.}%
	\label{fig:blackfriday:all}%
\end{figure}

\begin{figure}%
	\centering%
	\includegraphics[width=\linewidth]{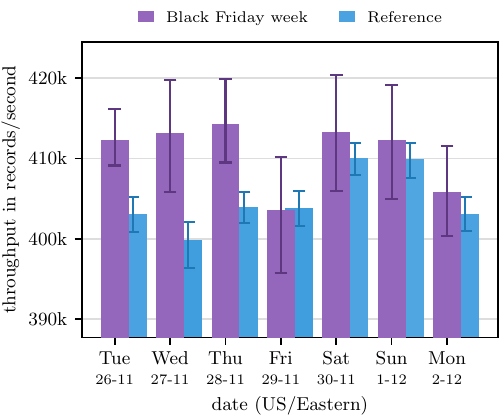}%
	\caption{Mean daily throughput around Black Friday compared to the daily pattern described in \cref{sec:patterns}.}%
	\label{fig:blackfriday:daily}%
\end{figure}

We repeated the periodic benchmark execution around Black Friday 2024.
\Cref{fig:blackfriday:all} shows the average throughput of each run, revealing a drop on Friday morning followed by recovery on Saturday morning.
To better highlight this effect, \cref{fig:blackfriday:daily} summarizes the average throughput per day and contrasts it with the baseline daily pattern identified in \cref{sec:patterns}.
The results show a clear dip in performance on the Friday compared to the three preceding and two following days, although the differences are not statistically significant due to overlapping confidence intervals.

Interestingly, the three days before Black Friday exhibit a statistically significantly throughput increase of $2.3\%$ to $3.3\%$ compared to the corresponding weekdays in the reference pattern.
In contrast, Black Friday itself shows no measurable deviation from the typical Friday performance.
For the days following Black Friday, we observe slightly higher performance, the effect is not statistically significant when compared to the reference pattern.

A possible explanation for these observations is the generally higher performance we measured in November 2024 compared to the summer months, followed by a temporary dip on Black Friday itself. Another explanation could be that the cloud provider proactively provisions additional computing resources in anticipation of Black Friday, which may elevate performance in the preceding days. In either case, our results suggest that Black Friday introduces a small but noticeable source of variability in cloud performance.

\section{Conclusions}

Our study confirms that application-level benchmark performance in the cloud exhibits noticeable variability. In contrast to related work, we identify clear effects of the time of day, the weekday, and global events such as Black Friday. Nevertheless, these effects are relatively small and only become relevant when targeting very small performance differences below 5\%.

\paragraph{Acknowledgments}
We would like to thank JKU and Dynatrace for co-funding this research.

\renewcommand*{\bibfont}{\footnotesize}
\renewbibmacro{in:}{}
\printbibliography

\end{document}